\documentclass[12pt]{article}

\usepackage{amsmath}
\usepackage{amssymb}
\usepackage{amsfonts}
\usepackage{epsfig}
\usepackage{float}
\usepackage[a4paper]{geometry}

\def\d{{\rm d}}
\def\i{{\rm i}}
\def\df#1#2{\frac{{\rm d}#1}{{\rm d}#2}}
\def\pd#1#2{\frac{\partial#1}{\partial#2}}

\def\H{{\cal H}}
\def\HH{\boldsymbol{\cal H}}

\def\M{{\cal M}}

\def\ave#1{\left\langle#1\right\rangle}

\def\ang#1#2{\left\langle#1,#2\right\rangle}
\def\comm#1#2{\left[#1,#2\right]}
\def\sch{Schr\"odinger}
\def\abs#1{\left|#1\right|}

\def\HHF{\hat{\H}_{\!F}}

\begin{document}



\title{A nonlinear quantum model of the Friedmann universe}

\author{Charles Wang\\
Department of Physics, Lancaster University\\
Lancaster LA1 4YB, UK\\
E-mail: c.wang@lancaster.ac.uk}

\date{}

\maketitle

\abstract{A discussion is given of the quantisation of a physical
system with finite degrees of freedom subject to a Hamiltonian
constraint by treating time as a constrained classical variable
interacting with an unconstrained quantum state. This leads to a
quantisation scheme that yields a \sch-type equation which is in
general nonlinear in evolution. Nevertheless it is compatible with
a probabilistic interpretation of quantum mechanics and in
particular the construction of a Hilbert space with a Euclidean
norm is possible. The new scheme is applied to the quantisation of
a  Friedmann universe with a massive scalar field whose dynamical
behaviour is investigated numerically.}

\vskip 7mm

\noindent
PACS numbers: 04.60.Ds, 04.60.Kz


\section{Introduction}

An elegant way of formulating classical particle dynamics is via
an action principle that is invariant under a general change of
time parameterisation. Such a formulation is called a
parameterised theory. In non-relativistic particle dynamics a
parametric description is obtained by regarding Newtonian time of
the particle as a dynamical variable depending on a newly
introduced ``parameter time''  \cite{lanczos}. As this choice is
arbitrary the corresponding Lagrangian admits a symmetry under
re-parameterisation of time, leading to a constraining relation
called the Hamiltonian constraint. Consequently not all canonical
variables (including the original Newtonian time and its conjugate
momentum) can be given arbitrary initial data. Rather, the
Hamiltonian constraint must be satisfied initially which will be
preserved under the canonical equations of motion. This is to be
expected since parameterising time cannot change physical
kinematics, the apparent extra degree of freedom associated with
time must be subject to a constraint.

In relativity the parameterised theory is ideally suited for
formulating particle dynamics where time is treated on
an equal footing with
other coordinates. This is especially the case for a
particle moving in a (pseudo) Riemannian geometry \cite{kuchar}.
The canonical description of a parameterised relativistic particle
is similar in structure to that of a non-relativistic particle.
However the constrained Hamiltonian is now quadratic in the
canonical momenta of both position and time coordinates. This is
due to the (general) covariance of relativistic kinematics and can be analysed
within the established methodology in classical theory. However this
appears to present a number of challenges for the quantisation of
a relativistic particle compatible with standard interpretations.
As is well-known, a straightforward application of canonical
quantisation results in a Klein-Gordon equation. Unfortunately its
solutions belong to a function space with an indefinite norm and
cannot be interpreted as wavefunctions describing a probability
amplitude in the usual manner. Although this difficulty can be
circumvented upon ``second quantisation'' in the context of
relativistic quantum field theory, theoretical interest remains in
search for a satisfactory direct quantisation of a relativistic
particle. One important reason for this is that a functional form
of relativistic particle dynamics naturally arises from canonical
general relativity \cite{DeWitt_1967}. Furthermore
systems with finite degrees of freedom subject to a constrained
Hamiltonian formally resembling that of relativistic particle
dynamics in curved spacetime often feature in various
cosmological models with or without coupled matter fields.  These
models are normally constructed by imposing high symmetry on
spacetime geometry and matter distribution in order to gain
insight into aspects of an underlying full theory. Note that in
these models the ``metric'' refers to the (reduced) ``supermetric'' (of
spacetime metric) and the ``relativistic particle'' represents a
collective gravitational and matter field point.

In \cite{Blyth_Isham_1975} a ``square-root Hamiltonian'' method was
investigated
in an attempt to overcome the problems mentioned with
the ``Klein-Gordon approach''. Given a choice of time coordinate
it is possible to construct a wave equation that is first order in
time and possesses a ``\sch{} structure''. While in some cases
this structure
guarantees the existence of a Hilbert space with  a Euclidean norm
it may suffer from other drawbacks including the violation of
Hermiticity of the Hamiltonian operator if the square-root of a
negative eigenvalue must be taken through a spectral analysis
\cite{Carlip_2001}.

Motivated by these circumstances an alternative canonical type of
quantisation for a physical system with finite degrees of
freedom subject to a Hamiltonian constraint is considered by
treating time as a constrained classical variable interacting with
an unconstrained quantum state. Following a brief review of the
basic parameterised theory of a non-relativistic particle in
curved space and its Dirac quantisation in section~\ref{cq}, a new
``parametric quantisation'' is introduced. For non-relativistic
particle dynamics, this new scheme is shown to reproduce results
based on Dirac quantisation. In section~\ref{pq}, parametric
quantisation of a relativistic particle in curved spacetime is
performed for a choice of time-slicing. In this case a ``nonlinear
\sch{} equation'' may be derived that is locally equivalent to the
full set of evolution equations from parametric quantisation under
certain conditions. In section~\ref{fu} the theory is applied in
the context of a simple Friedmann cosmological model with the
resulting equations analysed numerically. Concluding remarks are
made in section~\ref{cr} where some prospects for future work are
also discussed.

\section{Classical and quantum dynamics of a non-relativistic particle}
\label{cq}

Consider
the motion of a non-relativistic particle
subject to a potential
in a Riemannian $n$-space.
Denote by $\tau$ the
Newtonian time and introduce the
coordinates
$( q^\alpha )$  ($0 \le \alpha \le n$ with $q^0:=\tau$). The time-dependent spatial metric
components are denoted by
$\gamma_{ab} = \gamma_{ab}(q^\alpha) = \gamma_{ab}(q^c, \tau)$.
Here and throughout Greek letters such as $\alpha, \beta$ denote
spacetime indices with range $0,1,\cdots n$, whereas Latin letters such as $a, b$ denote
spatial indices with range $1,2,\cdots n$.
The particle motion is described by the
trajectory $( q^\alpha(t) ) = ( q^a(t), \tau(t) )$
parameterised by a parameter time $t$. The choice of $t$
is related to a positive function $N=N(t)$ that sets the parameterisation gauge
and enters into the generic form of the Hamiltonian
\begin{equation}\label{}
H(N, q^\alpha, p_\alpha) = N \H(q^\alpha, p_\alpha)
\end{equation}
that
generates the dynamics of the particle for some
$\H$.
For a non-relativistic particle
with canonical momenta $\varpi:=p_0$ and
$p_a$ conjugate to $\tau$ and $q^a$ respectively,
$\H$ is linear in $\varpi$ and quadratic in
$p_a$ as given by
\begin{equation}\label{nonrelH0}
\H = \varpi + \frac12\,\gamma^{a b} p_a p_b + V
\end{equation}
where $V(q^\alpha)=V(q^c, \tau)$ is a
time-varying potential and the particle's mass has been normalised to unity.
Based on an action principle
using the Lagrangian
\begin{equation}\label{}
L = \dot{q}^\alpha p_\alpha - H
\end{equation}
where $\dot{\;} = \df{}{t}$,
the equations of motion can be derived
in the following canonical form:
\begin{align}
\label{eqa}
  \df{{q}^a}{t} &= N\pd{\H}{p_a},\quad
  \df{{p}_a}{t} = -N\pd{\H}{q^a} \\
\label{eq0}
  \df{\tau}{t} &= N\pd{\H}{\varpi},\quad
  \df{\varpi}{t} = -N\pd{\H}{\tau}.
\end{align}
In addition the
Hamiltonian constraint
\begin{equation}\label{h0}
\H=0
\end{equation}
arises from
variations with respect to
$N$. Provided a set of initial conditions for the canonical variables
$q^\alpha$ and  $p_\alpha$ compatible with \eqref{h0} is given,
this constraint will be preserved by the evolution of
the canonical variables satisfying
\eqref{eqa} and \eqref{eq0} for a choice of (positive) $N(t)$.
Clearly the particle trajectory is invariant under
re-parameterisation:
$t \rightarrow t'$, $N \rightarrow N' := \df{t}{t'} N$.

A quantum theory of this particles may be established by applying
Dirac quantisation where, in the \sch{} picture, classical constraints are ``turned into''
operators that annihilate wavefunctions describing physical states.
Following this procedure, the operator
\begin{equation}\label{}
\hat{\H}_D(q^a, \hat{p}_a, \tau, \hat\varpi) :=
\hat{\varpi}
+
\frac12 \,\gamma^{-\frac14} \hat{p}_a \gamma^{\frac12} \gamma^{a b} \hat{p}_b \gamma^{-\frac14} + V
\end{equation}
where $\gamma := \det({\gamma_{ab}})$
is constructed by substituting
$p_\alpha \rightarrow \hat{p}_\alpha := -\i \frac{\partial}{\partial q^\alpha}$
in $\H$ and factor ordering so that the resulting term quadratic in $\hat{p}^a$ is
proportional to the  Laplacian operator on a
wavefunction $\Psi(q^a,\tau)$ of weight $\frac12$ with respect to the metric $\gamma_{ab}$.
Such a weighting choice is very useful in simplifying mathematical
expressions when dealing with the wavefunction of a quantum particle
moving in a time dependent spatial metric.

On Dirac quantisation the classical constraint \eqref{h0} becomes the
quantum constraint
\begin{equation}\label{WDW}
\hat{\H}_D\,\Psi = 0
\end{equation}
on any physical state $\Psi$.
Although \eqref{WDW} does not explicitly involve the parameter time $t$
it is natural to interpret $\tau$ as the evolution parameter for wavefunctions.
To see this more clearly we may rewrite \eqref{WDW} in the form
\begin{equation}\label{nonrelsch}
\i  \pd{}{\tau} \Psi
=
\hat{\HH}\Psi
\end{equation}
where
\begin{equation}\label{h}
\hat{\HH} := \frac12 \,\gamma^{-\frac14} \hat{p}_a \gamma^{\frac12} \gamma^{a b} \hat{p}_b \gamma^{-\frac14} + V
\end{equation}
plays the role of the ``true Hamiltonian'' as in conventional
\sch{} equations.

By treating $\tau$ as the evolution parameter the positive-definite  inner product
of two wavefunctions $\Psi_1$ and $\Psi_2$ at equal time $\tau$ is naturally
defined as
\begin{equation}
\label{angPsi}
\ang{\Psi_1}{\Psi_2} := \int  \!\Psi_1^* \Psi_2 \,\d^n\! q.
\end{equation}
Using this definition the operator $\hat{\HH}$
is self-adjoint and the standard probabilistic interpretation
of quantum mechanics applies. In particular if a wavefunction $\Psi$ is initially
normalised according to $\ang\Psi\Psi = 1$ this normalisation will be maintained
under \eqref{WDW} and therefore
the conservation of the total probability holds.

The derivation of the non-relativistic \sch{} equation \eqref{WDW}
above is often regarded as support for
Dirac quantisation. However it is worthy noting that
this derivation crucially relies on the fact that
$\H$ is {\em linear} in the momentum $\varpi$ conjugate to
the time variable $\tau$. For a relativistic particle, this is no longer
the case, since the Hamiltonian will be quadratic in all momenta
instead.

It is a purpose of this paper to point out the possibility
of deriving \eqref{WDW} based on an alternative quantisation scheme
in which the substitution
$\varpi \rightarrow -\i \frac{\partial}{\partial \tau}$
is {\em not} made and the explicit parameter time ($t$) dependence
will be retained. The guiding principle is that one of the dynamical variables
of a parameterised system should play the role of time. In the case of non-relativistic
particle dynamics, $\tau$ has been chosen to be this variable. As a time variable it is of
``classical nature'' but need not be treated merely as a parameter. Indeed in classical
parameterised theory such a time variable is treated as a dynamical
variable whose conjugate momentum is subject to a Hamiltonian constraint
without changing the underlying physical degrees of freedom.
In view of this we seek to formulate an analogous quantisation procedure
where the selected time variable is not quantised but treated as
a {\em constrained classical variable}.
The coupling of the classical time variable
with other quantised unconstrained variables
described by a wavefunction will be formulated subject to
certain requirements to validate essential physical interpretations.
These requirements include the existence of a
positive-definite norm for the Hilbert space of the wavefunctions
and the reduction to \eqref{eqa}, \eqref{eq0} and \eqref{h0} in the
classical limit.

To achieve this consider the operator
\begin{equation}
\label{nonrelH}
\hat{\H}(q^a, \hat{p}_a, \tau, \varpi)  := \varpi + \hat{\HH}
\end{equation}
where $\hat{\HH}$ takes the form as given in \eqref{h}.
The operator $\hat{\H}$ is constructed
in the way similar to that of $\hat{\H}_D$
with the substitution
$p_a \rightarrow \hat{p}_a = -\i \frac{\partial}{\partial q^a}$
in $\H$ but $\tau, \varpi$ now remain as classical functions of $t$.
In contrast to Dirac quantisation where $\hat{\H}_D$
is used in constructing a quantum constraint equation \eqref{WDW},
the operator $\hat{\H}$ will be used to formulate a \sch-type
equation
\begin{equation}
\i \pd{}{t}{\psi} = N \hat{\H} \psi
\label{nonrelweq}
\end{equation}
for a wavefunction $\psi(q^a,t)$ of weight $\frac12$
with respect to the metric $\gamma_{ab}$.
Nonetheless, like \eqref{angPsi},
the inner product of two wavefunctions $\psi_1$ and $\psi_2$
can be defined to be
\begin{equation}
\label{angpsi}
\ang{\psi_1}{\psi_2} := \int  \psi_1^* \psi_2 \, \d^n\! q
\end{equation}
under which the operator $\hat{\H}$ is self-adjoint.
It is readily seen that \eqref{nonrelweq} maintains the normalisation of
the wavefunction $\ang\psi\psi$ for {\em any} $\tau(t), \varpi(t)$.
This makes it possible to consistently impose the special normalisation $\ang\psi\psi = 1$ and
then interpret $\psi(q^a,t)$ as the probability
amplitude for measurements of the particle position $( q^a )$ on a spatial hypersurface
at time $\tau(t)$. For general normalisation, the expectation value
of an operator $\hat{O}=\hat{O}(q^a, \hat{p}_a, \tau, \varpi)$ may be defined as:
\begin{equation}\label{aveO}
\ave{\hat{O}} := \frac{\ang{\psi}{\hat{O}\,\psi}}{\ang{\psi}{\psi}}
\end{equation}
in a standard manner.
Sometimes it is useful to denote this expectation value by
$\ave{\hat{O}}_{\!\psi}$
to explicitly indicate that it is evaluated with respect to  the wavefunction $\psi$.
Using \eqref{nonrelweq} and \eqref{aveO},
the derivative of the expectation value $\ave{\hat{O}}$ with respect to the parameter time
$t$ can be shown to be
\begin{align}
\df{}{t}\ave{\hat{O}}
&=
%
%
\ave{\pd{\hat{O}}{\tau}}\df{\tau}{t}
+ \ave{\pd{\hat{O}}{\varpi}}\df{\varpi}{t}-\i N \ave{\comm{\hat{O}}{\hat{\H}}}
\label{dotO}
\end{align}
where $\comm{\;}{\;}$ denotes a commutator as usual.
The succinctness of this expression benefits from the use of
weight $\frac12$ wavefunctions and their operators.
Formulae \eqref{aveO} and \eqref{dotO} will now be used to construct
the governing equations for $\tau(t)$ and $\varpi(t)$.
Guided by canonical equations \eqref{eq0}
the equations
\begin{eqnarray}
  \df{\tau}{t} &=& N \ave{\pd{\hat{\H}}{\varpi}}, \quad  \df{{\varpi}}{t} = -N \ave{\pd{\hat{\H}}{\tau}}
\label{dott}
\end{eqnarray}
are adopted in a ``semi-classical'' fashion
to allow the coupling between variables $\tau, \varpi$ and wavefunction $\psi$
such that the expected classical limit may be recovered.
It can now be observed that, using \eqref{dotO}, $\ave{\hat{\H}}$ is a constant of motion
under \eqref{nonrelweq} and \eqref{dott}. That is,
\begin{eqnarray}
\df{}{t}\ave{\hat{\H}} &=& 0.
\end{eqnarray}
Therefore the condition
\begin{eqnarray}
\ave{\hat{\H}} &=& 0
\label{aveH}
\end{eqnarray}
may be consistently imposed
and regarded as the quantum analogue of the classical Hamiltonian
constraint \eqref{h0}. This condition may be considered  as a restriction on the initial classical variables
$\tau$ and $\varpi$ for any given arbitrary initial wavefunction $\psi$.

For a non-relativistic particle the quantum description using
\eqref{nonrelweq}, \eqref{dott} and \eqref{aveH} is in fact
equivalent to the single \sch-type equation \eqref{nonrelsch}.
To see this first note that in this case $\dot{\tau} = N$
by using \eqref{nonrelH} and \eqref{dott}.
Adopt the gauge $N=1$ so that we may set $\tau = t$. With this choice it is clear at once that
$\psi(q^a,\tau)$ is unitarily related to
$\Psi(q^a,\tau)$ satisfying \eqref{nonrelsch} via
\begin{equation}\label{}
\Psi(q^a,\tau) = \exp\left( \i\!\! \int^\tau\!\!\! \varpi(\tau') \d \tau' \right) \psi(q^a,\tau).
\end{equation}

Given a generic classical parameterised theory,
the procedure leading to equations \eqref{nonrelweq}, \eqref{dott} and \eqref{aveH}
will be referred to as ``parametric quantisation''.
Although both Dirac quantisation and parametric quantisation
give rise to equivalent results for non-relativistic particle dynamics,
these two schemes are in general inequivalent, as will be demonstrated
in the next section.

\section{Parametric quantisation of a relativistic particle}
\label{pq}


Let us now consider the motion of a particle in an $(n+1)$-dimensional pseudo-Riemannian
manifold $\M$ \cite{benn_tucker} with coordinates $(q^\alpha)$, $(0 \le \alpha \le n)$, in which the metric
has components $\gamma_{\alpha\beta}=\gamma_{\alpha\beta}(q^\mu)$ with signature $(-,+,+,\cdots)$.
The classical Hamiltonian for this particle
$H(N, q^\alpha, p_\alpha)= N \H$
with
trajectory coordinates $q^\alpha(t)$ and canonical momenta $p_\alpha(t)$
parameterised by $t$ associated with a positive gauge function $N(t)$
subject to a potential $V(q^\mu)$
is specified by
\begin{equation}\label{relHH}
\H(q^\alpha, p_\alpha ) = \frac12\, \gamma^{\alpha \beta} p_\alpha p_\beta + V.
\end{equation}
In order to follow closely the methods in the preceding section, the
manifold $\M$ is assumed to admit a
global time foliation so that
$(q^\alpha)$ forms
a single Gaussian coordinate chart \cite{kuchar_torre_1991a}
in which
\begin{equation}\label{}
\gamma_{00} = -1,\quad \gamma_{0a} = 0.
\end{equation}
With such a time-slicing
we denote by $\tau(t) := q^0(t)$ and
$
\gamma := \det({\gamma_{ab}})
$
where $1 \le a,b \le n$
as in the non-relativistic particle case. Thus
\eqref{relHH} becomes
\begin{equation}\label{relH}
\H = -\frac12\, \varpi^2 + \frac12\,\gamma^{a b} p_a p_b + V
\end{equation}
which generates
canonical equations of motion
\begin{align}
  \df{{q}^a}{t} &= N\pd{\H}{p_a},\quad
  \df{{p}_a}{t} = -N\pd{\H}{q^a} \\
  \df{\tau}{t} &= N\pd{\H}{\varpi},\quad
  \df{\varpi}{t} = -N\pd{\H}{\tau}
\end{align}
subject to the Hamiltonian constraint
\begin{equation}\label{relh0}
\H=0.
\end{equation}
Since $\H$ in \eqref{relH} resembles
\eqref{nonrelH0} we now
introduce
 the operator
\begin{equation}\label{relhatH}
\hat{\H} := -\frac12\, \varpi^2 + \hat{\HH}
\end{equation}
where $\hat{\HH}$ takes the form of \eqref{h}
and is self-adjoint with respect to the inner product
\begin{equation}\label{relinn}
\ang{\psi_1}{\psi_2} := \int  \psi_1^* \psi_2 \, \d^n\! q
\end{equation}
of any two wavefunctions $\psi_k = \psi_k(q^a,t)$
(of weight $\frac12$) with respect to the spatial metric $\gamma_{ab}$.
As in  the preceding section,
parametric quantisation now yields
the evolution equations for the wavefunction $\psi(q^a,t)$ and
constrained classical variables
$\tau(t), \varpi(t)$ as follows:
\begin{equation}\label{}
\i \pd{}{t}{\psi} = N \hat{\H} \psi
\label{relweq}
\end{equation}
\begin{eqnarray}
  \df{\tau}{t} &=& N \ave{\pd{\hat{\H}}{\varpi}}, \quad  \df{{\varpi}}{t} = -N \ave{\pd{\hat{\H}}{\tau}}
\label{reldott}
\end{eqnarray}
\begin{eqnarray}
\ave{\hat{\H}} &=& 0
\label{relaveH}
\end{eqnarray}
where the definition of expectation values follows from \eqref{aveO}.
As in the non-relativistic case, the normalisation $\ang\psi\psi = 1$ can be
consistently imposed and the interpretation of $\psi(q^a,t)$ as the probability
amplitude for measurements of the particle position $( q^a )$ on a spatial hypersurface
at time $\tau(t)$ can be made.

It is also possible (at least locally) to reduce \eqref{relweq},  \eqref{reldott} and \eqref{relaveH}
for the relativistic case
to a single wave equation.
From  \eqref{relH}, \eqref{reldott} and \eqref{relaveH}
it follows that $\dot{\tau} = -N \varpi$ where
\begin{equation}\label{p0sol}
\varpi = \pm \sqrt{2\ave{\hat{\HH}}}.
\end{equation}
Hence in regions where $\ave{\hat{\HH}}>0$ and
$\varpi < 0$
one can explicitly eliminate $\varpi$ and set $t = \tau$ by
choosing the gauge
\begin{equation}\label{relN}
N=-\frac1{\varpi}
\end{equation}
and introducing
\begin{equation}\label{}
\Psi(q^a,\tau) := \exp\left( \frac{\i}2\!\! \int^\tau\!\!\! \varpi(\tau') \d \tau' \right) \psi(q^a,\tau)
\end{equation}
where $\psi(q^a,\tau)$ satisfies \eqref{relweq} with \eqref{relN}.
Since $\ave{\hat{\HH}}_\psi = \ave{\hat{\HH}}_\Psi$,
\eqref{relweq} becomes
\begin{equation}
\i \pd{}{\tau}{\Psi} = \frac{\hat{\HH}\,\Psi}{\sqrt{2\ave{\hat{\HH}}_\Psi}}.
\label{nonlinweq}
\end{equation}
We are therefore led to a nonlinear integro-partial differential
equation describing the quantum evolution of a relativistic particle under a chosen
time-slicing. Although nonlinear in time, \eqref{nonlinweq} does not present
difficulties in the probabilistic interpretation of its solutions. It follows from
the scaling invariance of \eqref{nonlinweq},
i.e. if $\Psi$ is a solution to \eqref{nonlinweq} so is  $C \Psi$ for any
complex constant $C$ and hence both $\Psi$ and
$C \Psi$ represent the same physical state \cite{kibble_1978, Weinberg 1989a}.

\section{Friedmann universe with a massive scalar field}
\label{fu}

It is not hard to see that
the Gaussian coordinate condition  $\{\gamma_{00} = -1, \gamma_{0a} = 0\}$
used in the parametric quantisation of a relativistic particle
can be extended to the condition
$\{\gamma_{00} = \gamma_{00}(\tau), \gamma_{0a} = 0\}$
by  changing the coordinate time $\tau \rightarrow \int^\tau\!\!\sqrt{-\gamma_{00}(\tau')}\,\d \tau'$
for some negative $\gamma_{00}(\tau)$.
With this coordinate condition one can repeat the
derivations of
\eqref{relweq},  \eqref{reldott} and \eqref{relaveH}
using the operator
\begin{equation}\label{relhatH1}
\hat{\H} = \frac12\,\gamma_{00} \varpi^2 + \hat{\HH}
\end{equation}
with the same forms of $\hat\HH$ and
inner product $\ang{\:}{}$ as defined in \eqref{h} and \eqref{relinn} respectively.

This slightly generalised coordinate condition makes it easier
for the application of the parametric quantisation to a simple
cosmological model for $n=1$, where a
Friedmann universe filled with a massive scalar field is
analysed.\footnote{Units of $c = \hbar = 16 \pi G = 1$ are adopted.}
In terms of the lapse function $N(t)$
and scale factor $R(t)$ as functions of the
time coordinate $t$, the
Robertson-Walker metric reads
\begin{equation}\label{RW}
g = -N^2\d t^2 + R^2 \sigma
\end{equation}
where $\sigma$ is the standard metric on the homogeneous and
isotropic 3-space of constant curvature $K$, with $K=1,0,-1$
corresponding to the closed, flat and open cases respectively. The
dynamics of this geometry coupled to a uniformly distributed but
time-dependent scalar field $\phi(t)$ of mass $m$ is described by
the Lagrangian \cite{Blyth_Isham_1975}:
\begin{equation}\label{}
L = -\frac{6 R }{N}\dot{R}^2 + \frac{R^3}{2N}\dot{\phi}^2 - \frac{N{m}^2}{2}R^3\phi^2 + 6 N K R
\end{equation}
where $\dot{\;} = \df{}{t}$. The corresponding action $\int L \d t$
is manifestly invariant under time re-parameterisation:
$t \rightarrow t'$, $N \rightarrow N' := \df{t}{t'} N$.
The scale factor $R$ will now be chosen as the
(intrinsic) time variable. Accordingly we
denote by $q^0 = R, q^1 = \phi$ with the corresponding conjugate
momenta given by
\begin{align}\label{}
  p_0 &= \pd{L}{\dot{R}} =  -\frac{12 R }{N}\dot{R} =:  \Pi\\
  p_1 &= \pd{L}{\dot{\phi}} = \frac{R^3}{N}\dot{\phi} =: p
\end{align}
where symbols $\Pi$ and $p$ have been introduced for conciseness in the subsequent
discussions. It follows that
the Hamiltonian $H = \dot{q}^\alpha p_\alpha - L  = N \H$
where $\alpha=0, 1$ with $\H$ having the form of \eqref{relHH}
in terms of the nonzero metric components and  potential as follows:
\begin{align}\label{}
\gamma_{00} &= -12 R,\quad \gamma_{11} = R^3 \\
V &= \frac{{m}^2}{2}R^3\phi^2 - 6 K R.
\end{align}
By substituting $p \rightarrow \hat{p} := - \i \pd{}{\phi}$ into
$\H$ we obtain the operator
\begin{equation}\label{Kop}
\HHF :=  -\frac{1}{2R^3}\pd{^2}{\phi^2} +  \frac{  {m}^2}{2} R^3 \phi^2-\frac{\Pi^2}{24R}
-
6 K R
\end{equation}
which is to be applied to a wavefunction $\psi(\phi,t)$ (of weight $\frac12$
with respect to the 1-dimensional metric $\gamma_{11} = R^3$)
whose norm at any time $t$ is given by
\begin{equation}\label{}
{\ang{\psi}{\psi}} = \int_{-\infty}^\infty \!\!\abs{\psi(\phi,t)}^2 \d\phi.
\end{equation}
For simplicity the normalisation condition
\begin{equation}\label{norm}
  {\ang{\psi}{\psi}} = 1
\end{equation}
will be imposed
on the wavefunction $\psi$. On parametric quantisation
the evolution equations for the wavefunction $\psi$ and
constrained classical variables $R(t)$ and $\Pi(t)$ are then
formulated in accordance with \eqref{relweq},  \eqref{reldott} and \eqref{relaveH}
as follows:
\begin{equation}\label{sch}
\i \pd{}{t} {\psi} = N \HHF {\psi}
\end{equation}
\begin{align}\label{dR1}
\frac1N  \df{R}{t} &= \ave{\pd{\HHF}{\Pi}}
= -\frac{\Pi}{12R}
\\
\frac1N  \df{\Pi}{t} &= - \ave{\pd{\HHF}{R}}
=
-\ave{\frac{3}{2R^4}\pd{^2}{\phi^2} +  \frac{3 {m}^2}{2} R^2 \phi^2}
-\frac{\Pi^2}{24 R^2} + 6 K
\end{align}
\begin{equation}\label{H0}
  \ave{\HHF} = 0.
\end{equation}

To proceed a (time-dependent) functional basis
for $\psi$ will be chosen in order to
reduce the above
integro-partial differential
equations into a system of
nonlinear ordinary differential equations
of infinite dimensions. This system may be
truncated to facilitate numerical simulation.
To this end consider
\begin{equation}\label{psin}
\psi_k(\phi,R) := ({m} R^3)^{1/4}\,\xi_k(\sqrt{{m} R^3}\,\phi)
\end{equation}
for integers $k \ge 0$ where
\begin{equation}\label{}
\xi_k(x) := 2^{-k/2} \pi^{-1/4} (k!)^{-1/2}e^{-x^2/2}\,H_k(x)
\end{equation}
in terms of the Hermite polynomials $H_k(x)$.
The functions $\psi_k$
satisfy the eigen equations
\begin{equation}\label{sch_har_tindep}
\left(-\frac{1}{2R^3}\pd{^2}{\phi^2} +  \frac{  {m}^2}{2} R^3 \phi^2\right)
\psi_k
=
{m}\left(k+ \frac 1 2\right)\psi_k
\end{equation}
and orthonormality
\begin{equation}\label{orthon}
\int_{-\infty}^\infty \!\!\psi_j(\phi,R)\psi_k(\phi,R) \d\phi = \delta_{jk}
\end{equation}
for any positive $R$. They are used to expand the normalised wavefunction:
\begin{equation}\label{psian}
\psi(\phi,t) = \sum_{k=0}^{\infty} c_k(t)\psi_k(\phi,R(t))
\end{equation}
with complex coefficients $c_k(t)$ satisfying $\sum_{k=0}^\infty  \abs{c_k}^2 = 1$.
The time-dependence of these coefficients is to be determined by dynamics.
Using the expansion \eqref{psian} it follows that
\begin{align}\label{ipsit}
\i\pd{\psi}t
&=
\left[
\i\dot{c}_k
+
\frac{\i N \Pi}{16 R^2}\left(
\sqrt{k (k-1)}\,c_{k-2}
-\sqrt{(k+2)(k+1)}\,c_{k+2}
\right)\right] \psi_{k}
\end{align}
and
\begin{equation}\label{HHFpsi}
\HHF
{\psi}
=
\sum_{k=0}^\infty \left[{m}\left(k+ \frac 1 2\right)
-\frac{\Pi^2}{24R}
-6 K R
\right]c_k\psi_k.
\end{equation}
With the aide of \eqref{sch_har_tindep}, \eqref{orthon}, \eqref{ipsit} and \eqref{HHFpsi},
we obtain
the following coupled differential equations for
$c_k(t)\; (k \ge 0)$, $R(t)$ and $\Pi(t)$
by
substituting \eqref{psian} into \eqref{sch}--\eqref{H0}:
\begin{align}
\label{eqck}
\frac{\dot{c}_k}N
&=
\i \left[
\frac{\Pi^2}{24R}
+6 K R
-\left(k+ \frac 1 2\right){m}
\right]c_k
+
\frac{\Pi}{16 R^2}\left[
\sqrt{(k+2)(k+1)}\,c_{k+2}
-
\sqrt{k (k-1)}\,c_{k-2}
\right]
\\
\label{eqR}
\frac{\dot{R}}{N}
&=
-\frac{\Pi}{12R}
\\
\label{eqPi}
\frac{\dot{\Pi}}{N} &= -\sum_{k=0}^\infty
\frac{3{m}}{R}\sqrt{(k+1)(k+2)}\,\Re (c_k^* c_{k+2})
-\frac{\Pi^2}{24 R^2} + 6 K
\end{align}
together with the constraint
\begin{equation}\label{hc}
\sum_{k=0}^\infty {m}\left(k+ \frac 1 2\right)\abs{c_k}^2
-\frac{\Pi^2}{24R}
-6 K R = 0
\end{equation}
derived from \eqref{H0},
and the normalisation condition
\begin{equation}\label{nc}
\sum_{k=0}^\infty \abs{c_k}^2 = 1.
\end{equation}

Based on these equations numerical simulations are performed for
different choices of the curvature parameter $K$ to explore the
dynamical behaviour of the model.
With the mass of the scalar
$m=10$ and gauge $N=1$ the initial scale factor $R(t)$ at $t=0$ is
set to $0.5$. The initial wave profile $\psi(\phi,0)$ is chosen to
be gaussian by setting $c_0(0)=1$ with $c_k(0)=0$ for $k\ge 1$,
corresponding to an initially ``ground state''-like wavefunction
compatible with the normalisation condition \eqref{nc}.
The initial value for the momentum $\Pi(0)$ is then determined by
solving the constraint \eqref{H0} at $t=0$ with  $c_k(0), R(0)$
substituted with their initial values above. This yields two signs
for $\Pi(0)$ equivalent to integrations forward or backward in
time $t$ (with a negative or positive $\Pi(0)$ respectively). With
these initial data and parameters, a truncated $k$ index range
$0 \le k \le 60$ is found to produce numerical results with
satisfactory accuracy.

\begin{figure}[H]
\begin{center}
\begin{picture}(8,6)(0,0)
\put(0,6.5){\epsfig{file=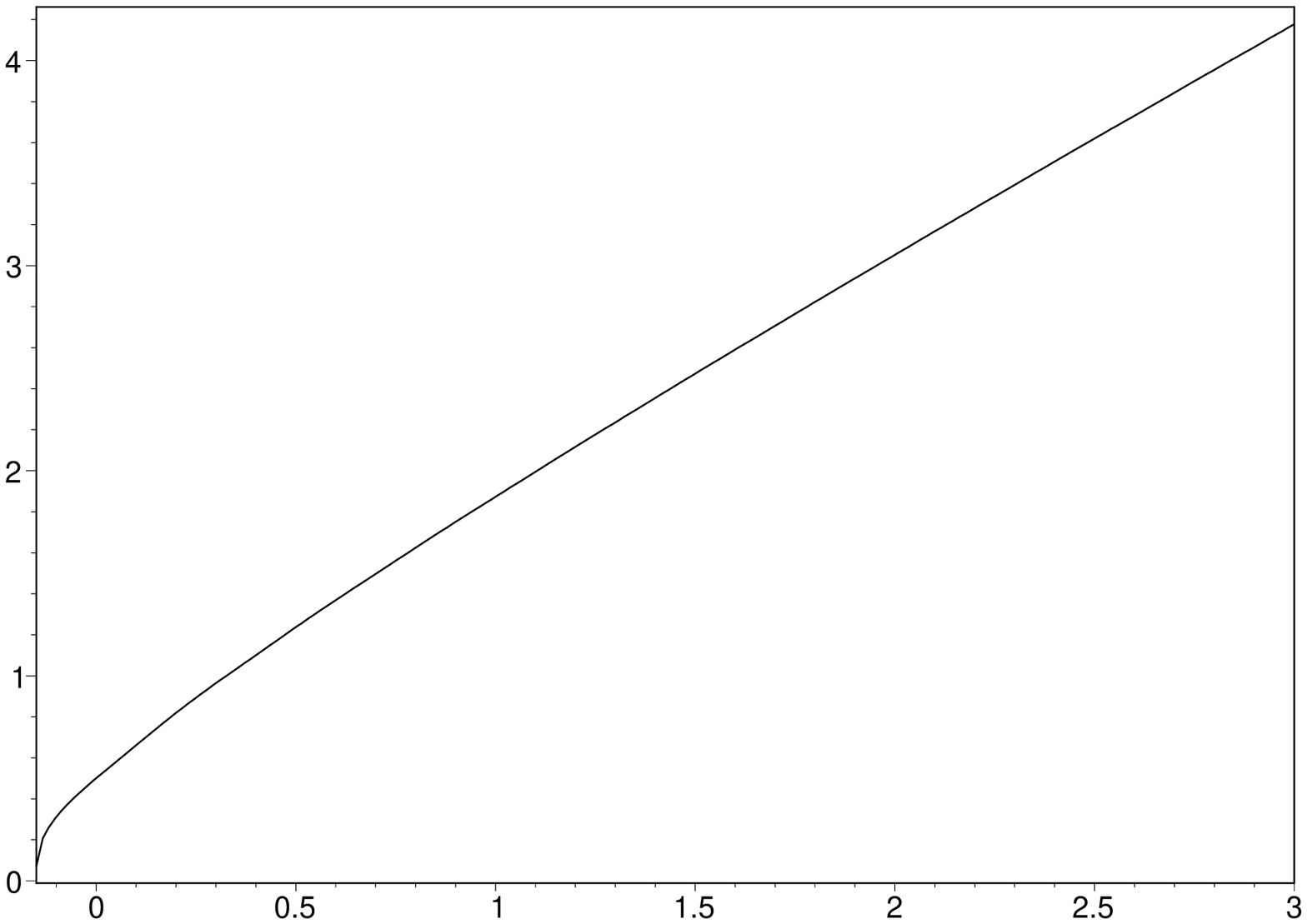,height=8cm,angle=270}}
\put(4,0.5){$t$}
\put(0,3.6){$R$}
\end{picture}
\caption{
Simulated $R(t)$ for an open Friedmann universe ($K=-1$).
An expanding universe with the
scale factor $R$  growing in $t$ is
clearly exhibited. This function approaches zero by decreasing $t$
where the Robertson-Walker spacetime becomes singular.}
\label{R_k-1}
\end{center}
\end{figure}

\begin{figure}[H]
\begin{center}
\begin{picture}(8,6)(0,0)
\put(0,6.5){\epsfig{file=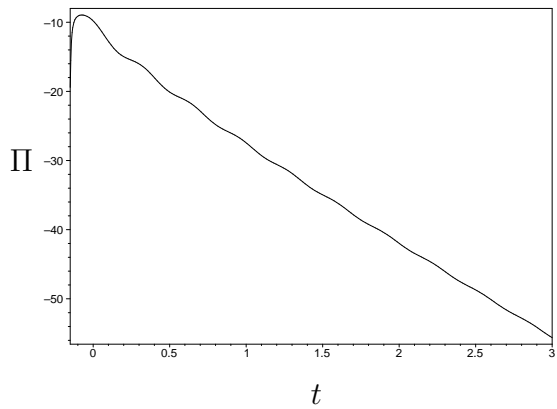,height=8cm,angle=270}}
\put(4,0.5){$t$}
\put(0,3.6){$\Pi$}
\end{picture}
\caption{
Simulated $\Pi(t)$ for an open Friedmann universe ($K=-1$).
Since in this case $\dot{R}$ does not
change the sign of $\Pi(t)$ remains negative.}
\label{PI_k-1}
\end{center}
\end{figure}

\begin{figure}[H]
\begin{center}
\begin{picture}(8,6)(0,0)
\put(0,7){\epsfig{file=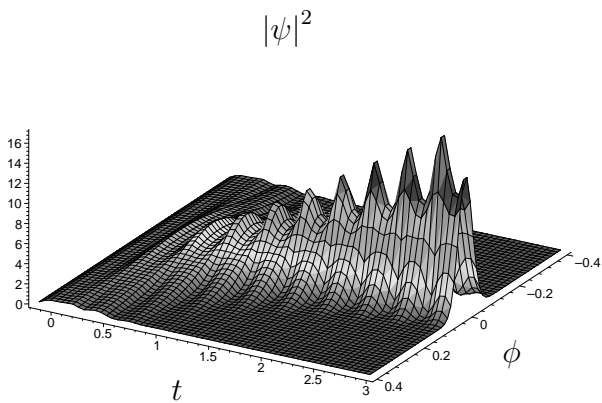,height=9cm,angle=270}}
\put(2.8,0.9){$t$}
\put(7.2,1.4){$\phi$}
\put(4,5.7){$\abs{\psi}^2$}
\end{picture}
\caption{Behaviour of $\abs{\psi(\phi,t)}^2$
based on simulated $\psi(\phi,t)$
for an open Friedmann universe ($K=-1)$.
This expression as a function of $\phi$ at a constant $t$ takes the shape of a single packet.
As the universe evolves
this packet becomes narrower and its height reduces to zero
as $t$ decreases to where $R\rightarrow0$.}
\label{rho_k-1}
\end{center}
\end{figure}

\begin{figure}[H]
\begin{center}
\begin{picture}(8,6)(0,0)
\put(0,7){\epsfig{file=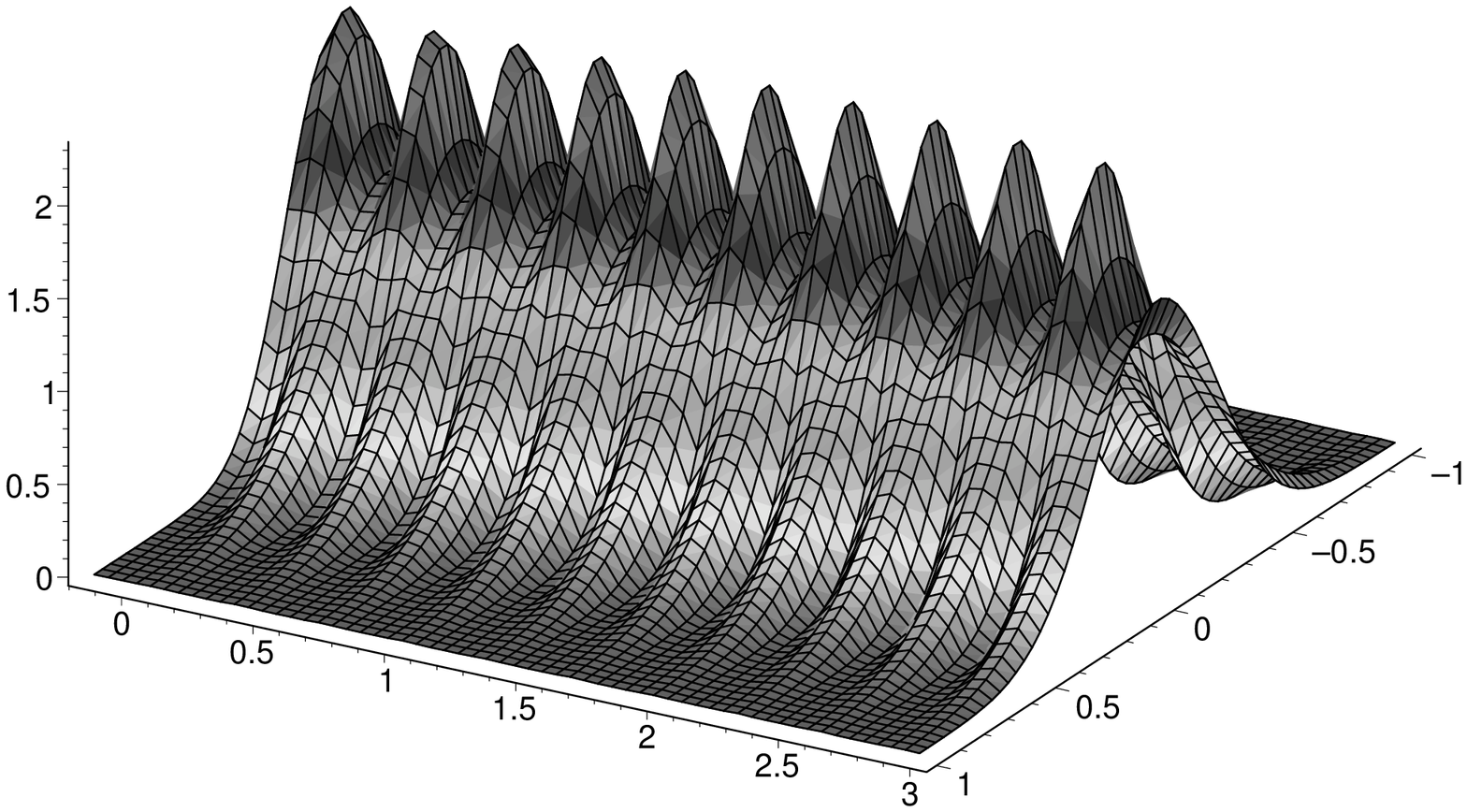,height=9cm,angle=270}}
\put(2.8,0.9){$t$}
\put(7.2,1.4){$R^{\frac32}\phi$}
\put(3.7,5.8){$\abs{R^{-\frac34}\psi}^2 $}
\end{picture}
\caption{
Behaviour of $\abs{R^{-\frac34}\psi}^2$
based on simulated $\psi(\phi,t)$ and $R(t)$
for an open Friedmann universe ($K=-1)$.
The profile of this expression as
a function of $R^{\frac32}\phi$ remains the shape of a single
packet whose width simply oscillates by either increasing $t$ or
decreasing $t$ even to where $R\rightarrow0$.}
\label{rho1_k-1}
\end{center}
\end{figure}

For an open Friedmann universe with $K=-1$ the simulated $R(t)$,
$\Pi(t)$ and $\abs{\psi(\phi,t)}^2$ are displayed in
figures~\ref{R_k-1}--\ref{rho_k-1} for $ -0.15 \le t \le 3.0$. The
wavefunction $\psi$ is evaluated using the expression
\eqref{psin} in terms of the numerical solutions for $c_k(t)$.
Figure~\ref{R_k-1} clearly exhibits an expanding universe with the
scale factor $R$  growing in $t$ and approaching a singularity
($R\rightarrow0$) with decreasing $t$. Since $\dot{R}$ does not
change the sign of $\Pi(t)$ remains negative as show in
figure~\ref{PI_k-1}. Figure~\ref{rho_k-1} shows that the
probability density $\abs{\psi}^2$ with respect to the measure
$\d\phi$ takes the shape of a single packet. As the universe evolves
this packet becomes narrower and its height seems to reduce to zero
as $t$ decreases to where $R\rightarrow0$. Despite this, the
alternative probability density $\abs{R^{-\frac34}\psi}^2$
(corresponding to the use of the weight 0 wavefunction
$R^{-\frac34}\psi$ with respect to the measure $R^{\frac32}\d\phi$)
exhibits a more steady behaviour in its evolution. As shown in
figure~\ref{rho1_k-1} the profile of $\abs{R^{-\frac34}\psi}^2$ as
a function of $R^{\frac32}\phi$ remains the shape of a single
packet whose width simply oscillates by either increasing $t$ or
decreasing $t$ even to where $R\rightarrow0$. The regular
behaviour of $R^{-\frac34}\psi$ suggests that the limiting value
of this function may serve to define the
initial quantum state of the Friedmann universe  as $R\rightarrow0$.

\begin{figure}[H]
\begin{center}
\begin{picture}(8,6)(0,0)
\put(0,6.5){\epsfig{file=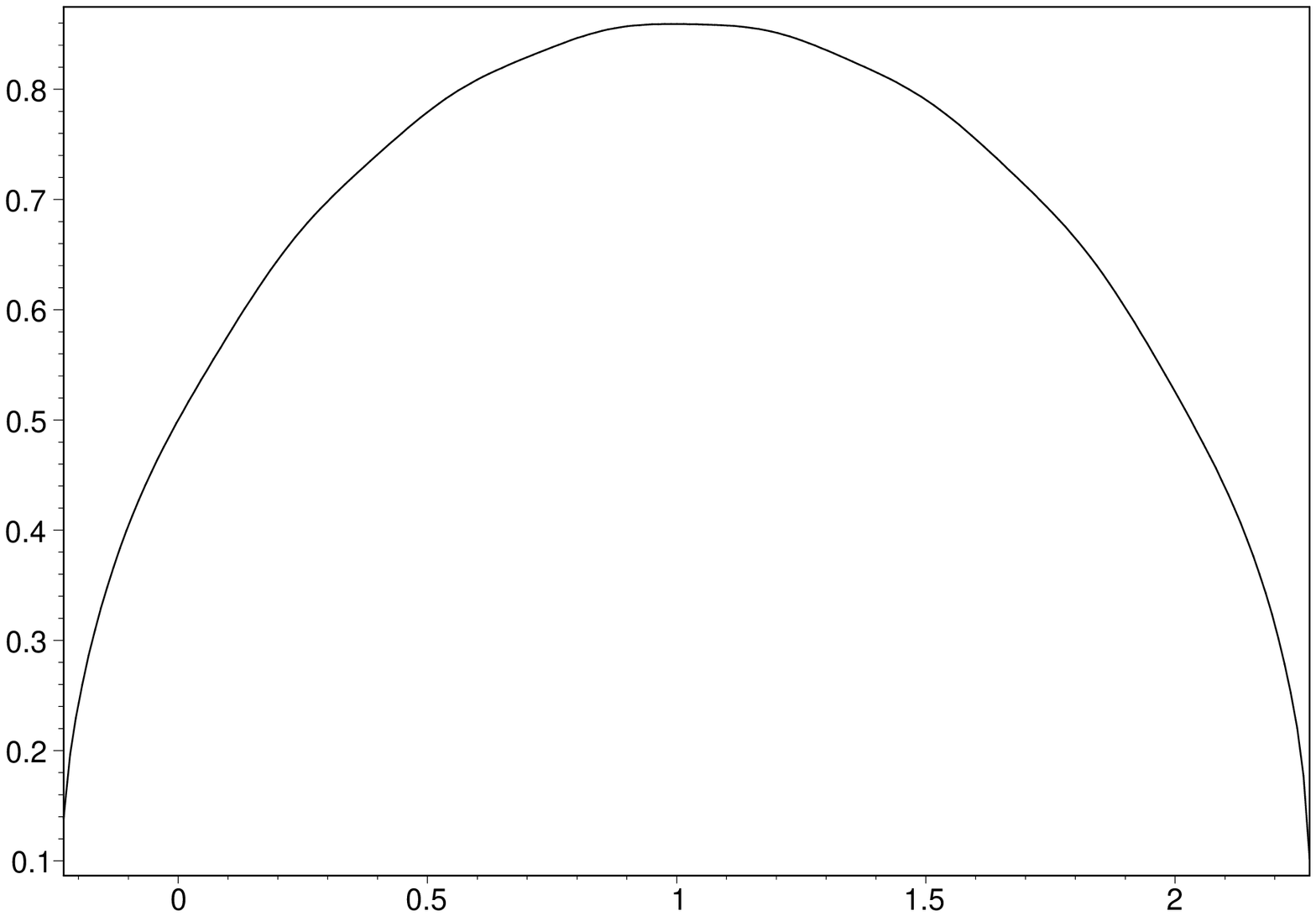,height=8cm,angle=270}}
\put(4,0.5){$t$}
\put(0,3.6){$R$}
\end{picture}
\caption{
Simulated $R(t)$ for a closed Friedmann universe ($K=1$).
Scenarios of Big Bang and Big Crunch as $R\rightarrow$ are
exhibited as in classical cases.}
\label{R_k1}
\end{center}
\end{figure}

\begin{figure}[H]
\begin{center}
\begin{picture}(8,6)(0,0)
\put(0,6.5){\epsfig{file=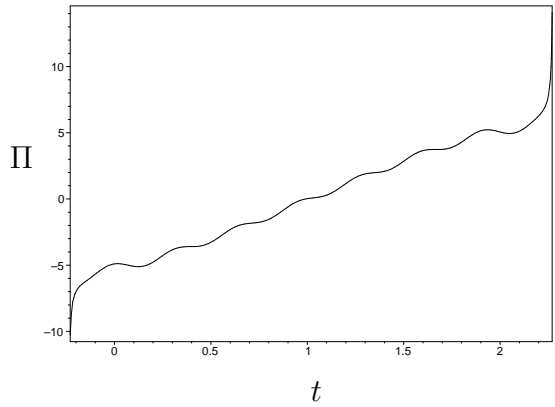,height=8cm,angle=270}}
\put(4,0.5){$t$}
\put(0,3.6){$\Pi$}
\end{picture}
\caption{
Simulated $\Pi(t)$ for a closed Friedmann universe ($K=1$)
This function changes
sign when the local maximum of $R$ takes place. This does
not affect the continuous evolution of the system.}
\label{PI_k1}
\end{center}
\end{figure}

\begin{figure}[H]
\begin{center}
\begin{picture}(8,6)(0,0)
\put(0,7){\epsfig{file=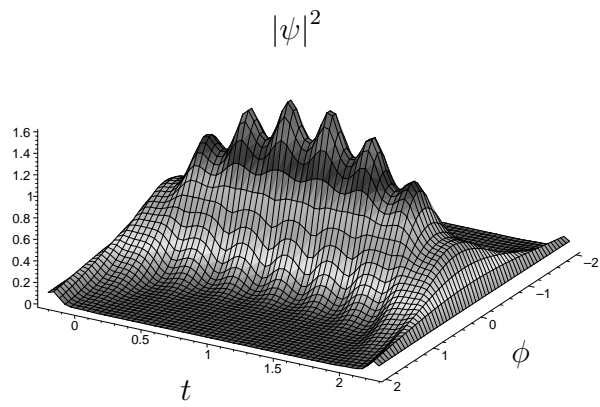,height=9cm,angle=270}}
\put(2.8,0.9){$t$}
\put(7.2,1.4){$\phi$}
\put(4,5.7){$\abs{\psi}^2$}
\end{picture}
\caption{
Behaviour of $\abs{\psi(\phi,t)}^2$
based on simulated $\psi(\phi,t)$
for a closed Friedmann universe ($K=1$).
This expression as a function of $\phi$ at a constant $t$ takes the shape of a single packet.
As
$t$ approaches the bounds of integration where $R\rightarrow0$ the
height of this packet also tends to zero.}
\label{rho_k1}
\end{center}
\end{figure}

\begin{figure}[H]
\begin{center}
\begin{picture}(8,6)(0,0)
\put(0,7){\epsfig{file=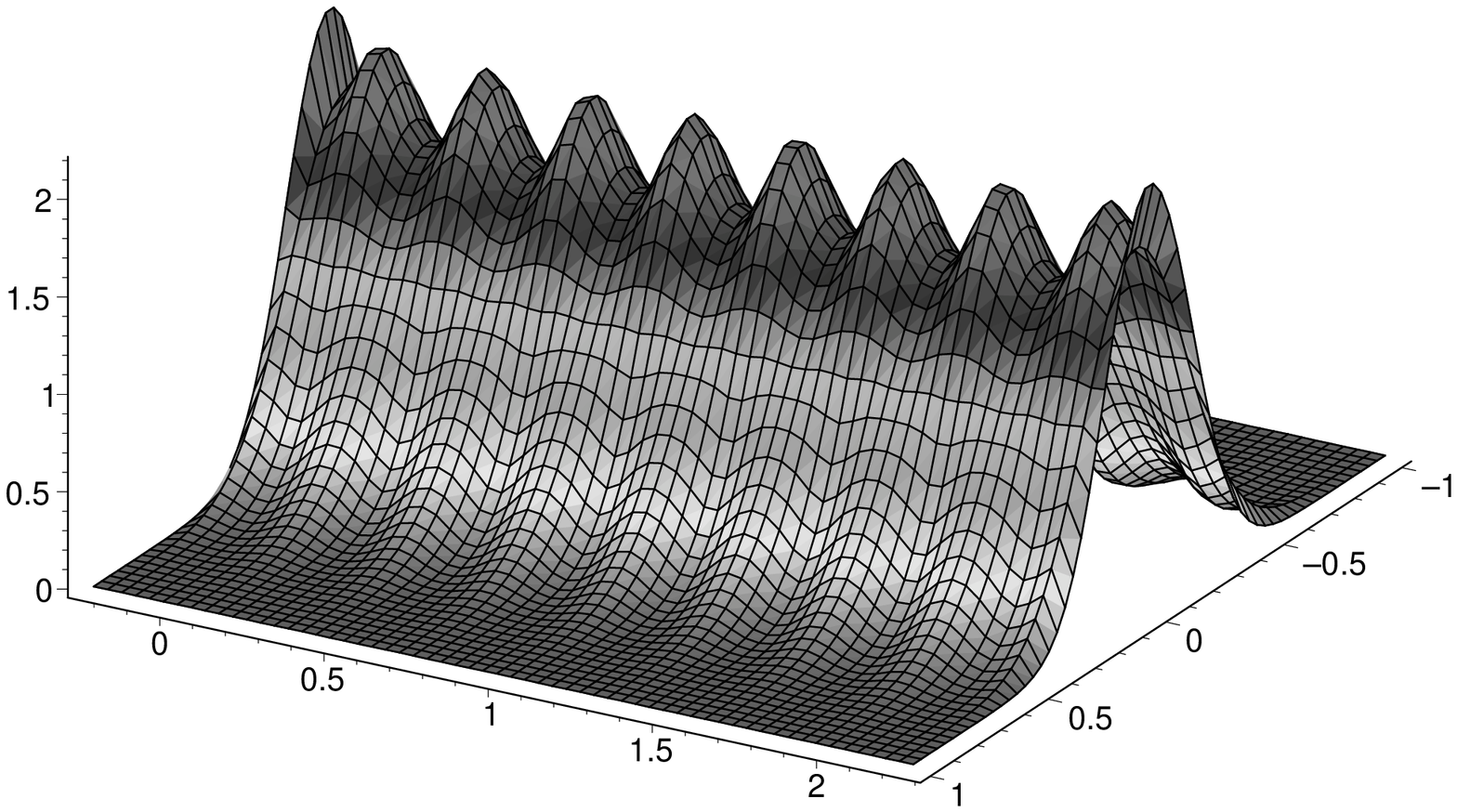,height=9cm,angle=270}}
\put(2.8,0.9){$t$}
\put(7.2,1.4){$R^{\frac32}\phi$}
\put(3.7,5.8){$\abs{R^{-\frac34}\psi}^2 $}
\end{picture}
\caption{Behaviour of $\abs{R^{-\frac34}\psi}^2$
based on simulated $\psi(\phi,t)$ and $R(t)$
for a closed Friedmann universe ($K=1$).
The profile of this expression as
a function of $R^{\frac32}\phi$ remains the shape of a single
packet whose width simply oscillates by either increasing $t$ or
decreasing $t$ even to where $R\rightarrow0$.}
\label{rho1_k1}
\end{center}
\end{figure}

Similar behaviours for $R$, $\Pi$ and $\psi$ can be observed from
simulation results for a flat Friedmann universe with $K=0$. The
numerical solutions for $R(t)$, $\Pi(t)$ and
$\abs{\psi(\phi,t)}^2$ for a closed Friedmann universe with $K=1$
are displayed in figures~\ref{R_k1}--\ref{rho_k1} for
$ -0.23 \le t \le 2.27$. Singularities are encountered as $R\rightarrow0$ at
both bounds of the numerical integration as shown in
figure~\ref{R_k1}. Notably in figure~\ref{PI_k-1}, $\Pi$ changes
sign when the local maximum of $R$ takes place. Nevertheless this
does not cause any problem for the evolution of the system. As in
the $K=-1$ case the probability density $\abs{\psi}^2$ with
respect to the measure $\d\phi$ has the shape of a single packet. As
$t$ approaches the bounds of integration where $R\rightarrow0$ the
height of this packet also tends to zero (figure~\ref{rho_k1})
whereas $\abs{R^{-\frac34}\psi}^2$ as a function of
$R^{\frac32}\phi$ remains the shape of a single packet with a steadily
oscillating width throughout the integration domain as shown in
figure~\ref{rho1_k1}. This provides further support for the use the
 limiting value of $R^{-\frac34}\psi$ as a function of
$R^{\frac32}\phi$ to identify the initial (or final)
quantum state of the Friedmann universe as $R\rightarrow0$.

\section{Concluding remarks}
\label{cr}

A quantisation scheme for a physical system subject to a
Hamiltonian constraint is presented based on canonical quantisation, nonlinear
quantum mechanics, couplings between classical and quantum
variables and analysis of the true degrees of freedom in
geometrodynamics \cite{Carlip_2001}. Motivated by the need for a
clarified role played by ``time'' in the ongoing efforts to
ultimately quantise gravity, the idea of treating time as a
constrained classical variable is put forward in this paper.
Instead of regarding time as an evolution parameter, the proposed
idea treats it as a dynamical variable while maintaining its
classical nature. It does not follow that the physical degrees of
freedom will increase by giving time a dynamical status since the
time variable and its canonical momentum are subject to the
(quantised) Hamiltonian constraint. This does imply that if the
Hamiltonian constraint only restricts the choice of classical
time and its canonical momentum, then the quantum degrees of
freedom can be regarded as unconstrained, thereby enabling much of the
standard quantisation technique to continue to apply. Of course the issue of the
preferred choice of time remains just as in the square-root
Hamiltonian approach. However it is worth emphasising that the
proposed parametric quantisation method is free from problems
associated with taking square roots of negative eigenvalues arising
from a spectral analysis. Therefore the choice of a time
coordinate is less restrictive for parametric quantisation.
Whether there exists a naturally preferred time variable given a
classical parameterised theory or possibility of establishing
equivalence for different choices of time within the framework of
parametric quantisation stays unaddressed in this paper and will
be considered in a later publication.
Finally in extending the
present work to the quantisation of gravity it is conceivable that
one of the traditional functional degrees of freedom in
geometrodynamics should be treated as a constrained classical
variable within the framework set up by this paper. York's
``extrinsic time'' \cite{York 1972} is amongst potential
candidates for such a treatment. In this respect it is of
interest to compare with the approach by
\cite{Kheyfets Miller 2000}.
The progress of these further researches will be reported
elsewhere.

\section{Acknowledgements}

The author is indebted to Professor Robin W Tucker for stimulating discussions
on this work
and is grateful to Professors Chris Isham and Malcolm MacCallum
for helpful comments on the final draft of this paper.
The research is supported in part by EPSRC and BAE Systems.

\newpage

\newpage

\newpage

\newpage

\end{document}